\documentclass[showpacs, amsmath, amssymb, amsthm, aps, jmp, thmsb, twocolumn]{revtex4}

\usepackage{color}
\usepackage{float}
\usepackage{amsmath}
\usepackage{amssymb}
\usepackage{amsfonts}
\usepackage{latexsym}
\usepackage{mathrsfs}
\usepackage[dvipdfm]{graphicx}

\newcommand{\be}{\begin{eqnarray}}
\newcommand{\ee}{\end{eqnarray}}
\def\({\left(}
\def\){\right)}

\newcommand{\bra}[1]{\langle #1 |}
\newcommand{\ket}[1]{| #1 \rangle}

\newcommand{\sla}[1]{\rlap{\kern .15em /}#1}

\def\jmp#1#2#3{{\it J.\ Math.\ Phys.}\ {\bf {#1}} ({#2}) #3}
\def\jpA#1#2#3{{\it J.\ Phys.}\ {\bf A{#1}} ({#2}) #3}

\def\prA#1#2#3{{\it Phys.\ Rev.}\ {\bf A{#1}} ({#2}) #3}

\def\prl#1#2#3{{\it Phys.\ Rev.\ Lett.}\ {\bf #1} ({#2}) #3}

\def\nat#1#2#3{{\it Nature} {\bf {#1}} ({#2}) #3}

\def\qmp#1#2#3{{\it Quant. Inf. Processing}\ {\bf {#1}} ({#2}) #3}
\def\qic#1#2#3{{\it Quant. Inf. Comput.}\ {\bf {#1}}({#2}) #3}

\begin{document}
\title{Entanglement Measures for Intermediate Separability of Quantum States}

\author{
Tsubasa Ichikawa,
Toshihiko Sasaki,
Izumi Tsutsui}

\affiliation{
High Energy Accelerator Research Organization (KEK),
Tsukuba, Ibaraki 305-0801, Japan
}

\begin{abstract}
We present a family of entanglement measures $R_m$ which 
act as indicators for separability of $n$-qubit quantum states into $m$ subsystems for arbitrary $2 \le m \le n$.    The measure $R_m$ vanishes if the state is separable into $m$ subsystems, and for $m = n$ it gives the Meyer-Wallach measure while for $m = 2$ it reduces, in effect, to the one introduced recently by Love {\it et al}.
The measures $R_m$ are evaluated explicitly for the GHZ state and the W state (and its modifications, the ${\rm W}_k$ or Dicke states) to show that these globally entangled states exhibit rather distinct behaviors under the measures, indicating the utility of the measures $R_m$ for characterizing globally entangled states as well.
\end{abstract}

\maketitle
\section{Introduction}

Quantum entanglement typifies one of the most striking aspects of quantum mechanics, posing profound questions on our commonsensical comprehension of the physical world.   The conceptual significance of entanglement 
was first pointed out 
in the celebrated EPR paper \cite{EPR}, where the nonlocal correlation of entangled states was regarded 
as a major obstacle for quantum mechanics to be a complete, realistic theory.   The validity of nonlocal reality was later examined by Bell \cite{Bellone, CHSH}, who put the conceptual problem to one which is testable in laboratory.
Since then, a variety of experiments have been conducted \cite{Aspect, Weihs, Rowe}, and by now we are almost convinced  that nonlocality does occur precisely as prescribed by quantum mechanics.   Although the nonlocal correlation generated by quantum entanglement cannot be used for communication \cite{Eberhard, GRW}, it suggests the existence of some nonlocal \lq influence\rq\ exerted between distant partners at a speed possibly exceeding that of light as reported by a recent experiment \cite{Salart}.  

In view of its salient characteristics, quantum entanglement is expected to play a vital role in our future technology
such as quantum computation and cryptography \cite{NC}.    Successful application of entanglement will in general require the ability of manipulating and measuring entangled $n$-qubit states at a reasonable level of accuracy.   Among them, characterization of entanglement is perhaps the most basic requisite, and for this there have been
a number of attempts including the use of canonical forms, entanglement witnesses and entanglement measures \cite{Horodecki, Pleno}.   These tools are certainly convenient for quantifying entanglement for a few small $n$ cases, but they
become almost intractable for large $n$ due to the exponential increase in the number of distinct structures allowed for the entangled states \cite{Duer}.   It seems, therefore,  inevitable that in order to quantify entanglement of generic $n$-qubit systems, we need to resort to some means specifically designed for the objectives to be achieved.

Among the many entanglement measures proposed so far \cite{Eisert, Yang, Horodecki, Pleno, Hassan}, the Meyer-Wallach (MW) measure \cite{MW} is notable in that it examines the {\it full separability}, {\it i.e.}, if the $n$-qubit state under inspection is a product state of all the $n$ constituent subsystems \cite{Brennen, Scott}.  
Recently, Love {\it et al.}~\cite{Love} proposed a measure which is \lq opposite\rq\ to the MW measure in the sense that it examines the {\it global entanglement}, {\it i.e.}, if the state admits no two subsystems
into which it can be decomposed as a product.
In the present paper, we present a family of entanglement measures $R_m$, $m = 2, 3, \ldots, n$ which 
can examine the {\it intermediate separability}, {\it i.e.}, if the state is a product state of arbitrary $m$ subsystems.  In particular, for $m = n$ our measure coincides with the MW measure, whereas for $m = 2$ it reduces, in effect, to the measure of \cite{Love}.   We show that, besides as indicators of intermediate separability, 
our measures can also be used in characterizing globally entangled states in general.  This is illustrated by the two standard globally entangled states, the GHZ state \cite{GHZ} and the W state \cite{Duer}  in $n$-qubit systems, which exhibit rather contrasting behaviors under our measures $R_m$ for various $m$.  
Analogous distinct behaviors can also be observed for the set of globally entangled ${\rm W}_k$ states (known as Dicke states), which are introduced as modified W states for $1 \le k \le \lfloor n/2 \rfloor$, with ${\rm W}_{n/2}$ furnishing the maximally entangled state for the MW measure $R_n$.   

The present paper is organized as follows.   In the next section, we introduce the family of entanglement measures with the required intermediate separability, and show that both the MW measure and the measure of \cite{Love} appear at the two ends of the set.   We then analyze, in section 3, the globally entangled GHZ and W states in terms of the measures introduced.   In section 4, the analysis is extended to the ${\rm W}_k$ states.  Section 5 is devoted to our conclusion and discussions.

\section{Entanglement measures as indicators of intermediate separability}

The system we consider is an $n$-qubit system whose quantum states are described by vectors in 
the Hilbert space $\mathbb{C}^{2^n}$.   In order to discuss its arbitrary subsystems, we label the $n$ constituent $1$-qubit systems by integers so that any subsystem consisting of some of the constituent systems is specified by a subset of ${\cal T} = \{1, 2,  \ldots, n\}$.   Let ${\cal P} = \{s_i\}_{i=1}^m$ be a partition of  ${\cal T}$, {\it i.e.},
\be
\bigcup_{i=1}^m s_i = {\cal T}
\quad
{\rm and}
\quad
s_i\cap s_j=\emptyset
\quad
{\rm for}
\quad
i\neq j.
\ee
Each subset $s_i$ determines a corresponding subsystem of the total system $\mathbb{C}^{2^n}$, and hence we may use $s_i$ to refer to the subsystem specified by the subset.   We denote by $\bar s_i$ the subset complementary to $s_i$ in ${\cal T}$ with $s_i\cup \bar s_i = {\cal T}$.

Now, given a pure state $\ket{\psi}$, let $\rho_{s_i}$ be the reduced density matrix in the subsystem $s_i$ obtained by taking the trace of   the density matrix $\rho = \ket{\psi}\bra{\psi}$ over the complementary space $\bar s_i$.   
Letting also $\vert s_i \vert$ be the number of elements (constituents) in the subset $s_i$,
we recall that the quantity
\be
\eta_{s_i} (\psi)= N(|s_i|)
  \( 1 - {\rm tr} \rho_{s_i}^2 \), 
\,\,\, N(|s_i|)
 =\frac{2^{\vert s_i \vert}}{2^{\vert s_i \vert}-1},
\label{meeta}
\ee
introduced in \cite{Love} vanishes
$\eta_{s_i} (\psi) = 0$ iff the state $\ket{\psi}$ is separable with respect to $s_i$ and $\bar s_i$.  Here, the normalization factor $N(|s_i|)$ in (\ref{meeta}) is chosen so that we have $\eta_{s_i} (\psi) = 1$ 
when the reduced state is maximally mixed $\rho_{s_i} = {1\over {2^{\vert s_i \vert}}} I$.   
This quantity $\eta_{s_i}$ is in fact a generalization of the (squared) concurrence \cite{Wootters} and can also be regarded as the quantum linear entropy \cite{BZ}.
From the quantities $\eta_{s_i}$ in (\ref{meeta})  obtained for all the subsets $s_i$ in ${\cal P}$, we evaluate the \lq average\rq\ value for the partition ${\cal P}$ by the arithmetic mean,
\be
\xi_{\cal P} (\psi)= {1\over m} \sum_{i=1}^m \eta_{s_i} (\psi).
\label{xsies}
\ee
Clearly, we have $\xi_{\cal P} (\psi) = 0$ iff the state $\ket{\psi}$ is separable according exactly to the partition ${\cal P}$ of the total set ${\cal T}$.

Out of all possible partitions ${\cal P}$ of ${\cal T}$, we may choose those ${\cal P}$ consisting of $m$ subsets for some $m$ in the range $2 \le m \le n$, and evaluate the geometric mean of the quantities $\xi_{\cal P} (\psi)$.  Namely, if $d({\cal P})$ is the number of subsets of the partition ${\cal P}$, 
we consider 
\be
R_m(\psi) := \left( \prod_{
\scriptstyle d({\cal P}) = m} \xi_{\cal P} (\psi) \right)^{1/{ S(n, m) }},
\label{rmeasures}
\ee
where
\be
S(n, m) 
= \sum_{k=1}^m {{(-1)^{m-k} k^{n-1} }\over{(k-1)!\, (m-k)!}}
\ee
is the Stirling number in the second kind \cite{Abramowitz}, which represents the number of all possible partitions of the integer $n$ into $m$ subsets, or the number of partitions ${\cal P}$ with $d({\cal P}) = m$.
The quantities $R_m(\psi)$ possess the important property:
\begin{equation}
R_m(\psi)= 0 \quad
\Leftrightarrow \quad
\begin{array}{ll}
\hbox{$\ket{\psi}$ is separable (at least)} \\
\hbox{ in $m$ subsystems in $\mathbb{C}^{2^n}$}.
\end{array}
\label{rmeasurestwo}
\end{equation}
Besides, since $R_m(\psi)$ are formed from $\eta_{s_i} (\psi)$ which are all entanglement monotones \cite{Love}, each of them, $R_m(\psi)$, $m = 2, \ldots, n$,  qualifies as an entanglement measure.   In particular, for $m = n$ where the subsets $s_i$, $i = 1, \ldots, n$, correspond to all the constituent subsystems, we have 
\be
R_n(\psi) = {1\over n} \sum_{i=1}^n 2 \left\{ 1 - {\rm tr} \rho_{s_i}^2 \right\},
\label{mwmeasure}
\ee
which is precisely 
the MW measure \cite{MW, Brennen} (see also \cite{Barnum04, Somma}).  On the other hand, at the other end
$m = 2$ we have the partitions $S = \{s_1 = s, s_2 = \bar s\}$.   Choosing the subset $s$ so that
$\vert s\vert \le \vert \bar s\vert$, and noting $S(n, 2) = 2^{n-1} - 1$, we find
\be
R_2(\psi) = \left( \mathop{{\prod}'}_{
\scriptstyle 1 \le \vert s\vert \le \vert \bar s\vert} c(s) \, \eta_{s} (\psi) \right)^{1/({ 2^{n-1} - 1 })},
\label{Rtwo}
\ee
where the prime on the product symbol indicates  that either one of the subsets $s$ and $\bar s$ is included when $\vert s\vert = \vert \bar s\vert$, and the coefficients $c(s)$ are given by
\be
c(s) = 1 - {1\over 2} \cdot {{2^{\vert \bar s\vert} - 2^{\vert s\vert}}\over{2^n - 2^{\vert s\vert}}}.
\ee
The measure $R_2(\psi)$ is equivalent to the measure proposed by Love {\it et al}.~\cite{Love}, apart from the factor $c(s)$ which varies between the maximum $c(s) = 1$ for $\vert s\vert = \vert \bar s\vert$ and the minimum $c(s) = 1/2 + 1/(1-2^{1-n}) > 1/2$ for $\vert s\vert = 1$.

\begin{figure}
\includegraphics[height=1.5in]{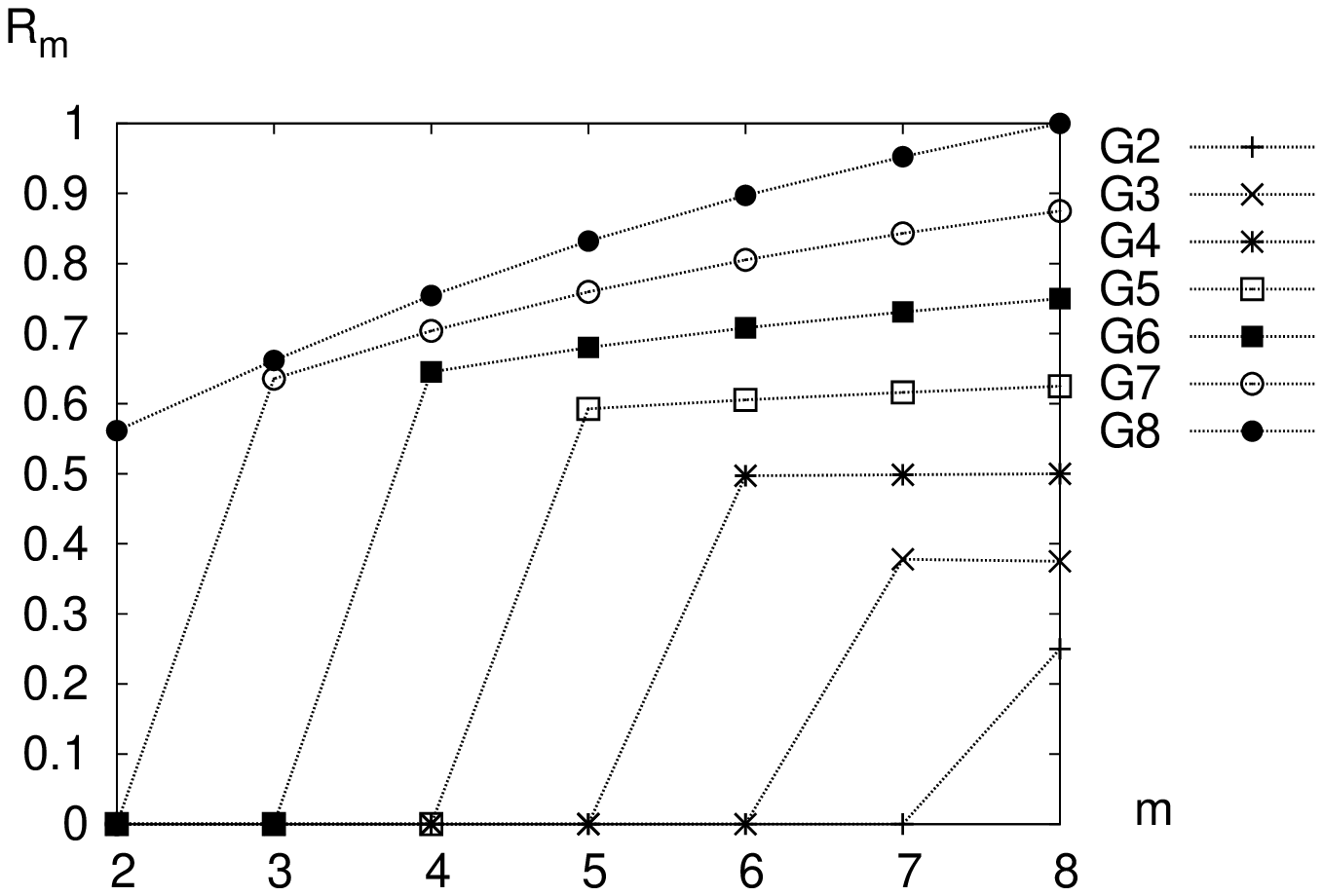}\\
\vspace*{4mm}\hspace*{2.4mm}
\includegraphics[height=1.5in]{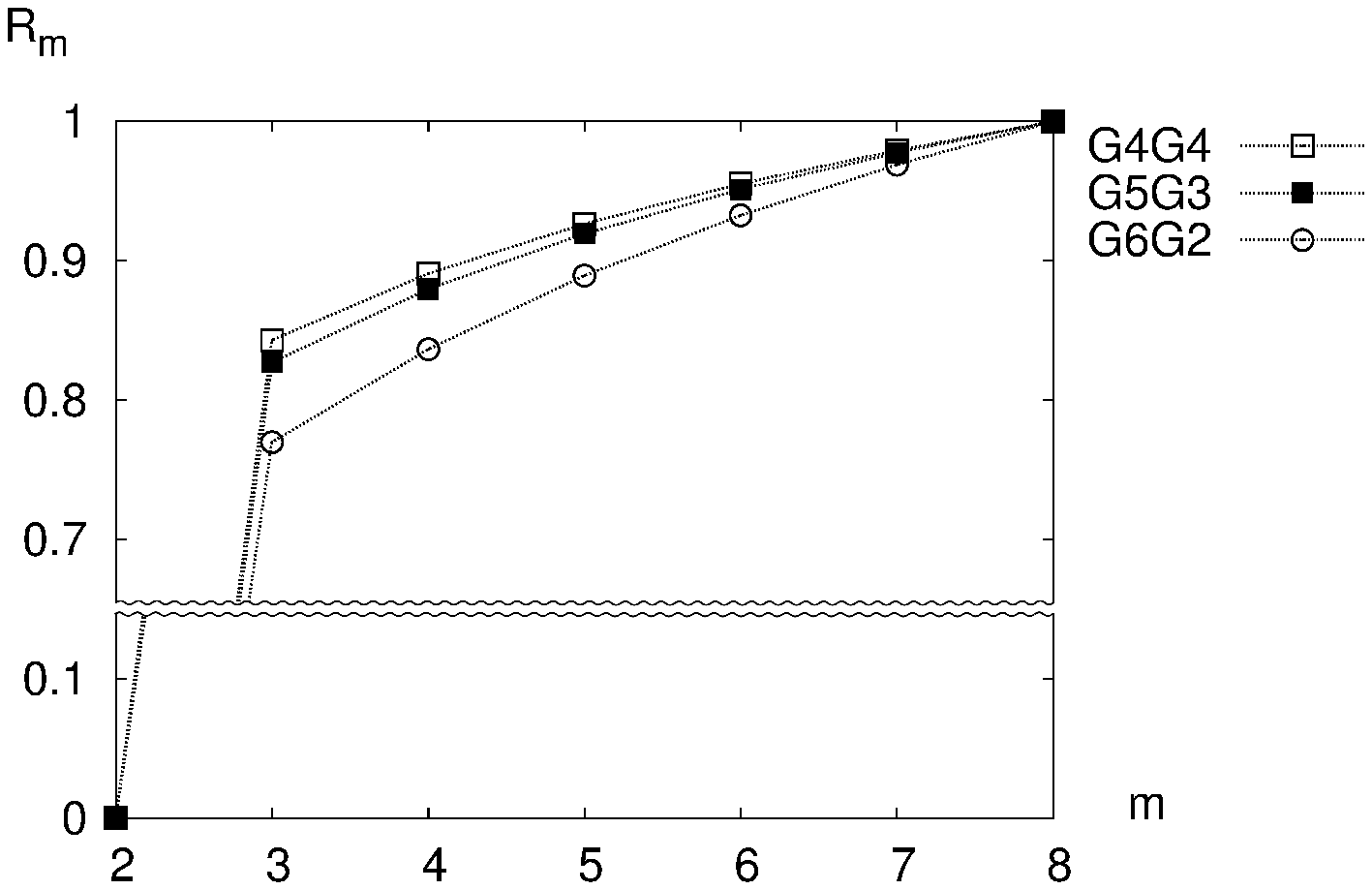}
\caption{The values of the measures $R_m$ evaluated for two sets of $8$-qubit states:  $|{\rm G}k \rangle |0\rangle^{\otimes(8-k)}$  for $k = 3, \ldots, 8$ (above) and $|{\rm G}k \rangle
|{\rm G}l \rangle$ with $(k,l) = (4,4), (5,3), (6,2)$ (below), which are represented by the tags \lq ${\rm G}k$\rq\ and \lq ${\rm G}k {\rm G}l$\rq, respectively.
The vanishing values confirm the number of subsystems into which the states $|{\rm G}k \rangle |0\rangle^{\otimes(n-k)}$ are factorized.   All of the states $|{\rm G}k \rangle
|{\rm G}l \rangle$ which are separable only into two subsystems share the same values
$R_2$ = 0 and $R_8$ = 1 but allow different values for other $R_m$.}
\label{figone}
\end{figure}

For illustration, we consider, for example, the $n$-qubit state $|{\rm G}k\rangle |0\rangle^{\otimes(n-k)}$ 
with $|{\rm G}k \rangle$ being the $k$-qubit version of the GHZ state
\be
|{\rm GHZ}\rangle=\frac{1}{\sqrt{2}}\(|11\cdots 1\rangle+|00\cdots 0\rangle\).
\label{ghzstate}
\ee
The state $|{\rm G}k\rangle |0\rangle^{\otimes(n-k)}$ is clearly 
separable into $n-k + 1$ subsystems, and Figure \ref{figone} shows that the measures $R_m(\psi)$ for $k = 3, \ldots 8$ and $n=8$ indeed vanish at $m = 8-k + 1$, where we further observe that $R_m$ behave rather distinctively depending on the size $k$ of the entangled subsystem.   
The measures are also evaluated for the product states  $|{\rm G}k \rangle |{\rm G}l \rangle$ consisting of two GHZ states for $(k,l) = (4,4), (5,3), (6,2)$, and the result shows that $R_m$ can distinguish these states which are all indistinguishable under both of the MW measure (since $R_8$ = 1) and the measure by Love {\it et al}.~(since $R_2$ = 0).   These observations suggest that our measures $R_m(\psi)$, as a whole, may also be useful to 
characterize multipartite entanglement of the state $\ket{\psi}$ in addition to examining the intermediate separability.  This possibility is explored further in terms of the GHZ state and the W state later, where we also present an explicit procedure to evaluate the measures for these particular states. 

At this point, we mention that the quantity 
$\eta_{s_i} $ in (\ref{meeta}), which 
is an entanglement measure for the separability of the subset $s_i$, can be  
considered as the purity measure based on the subalgebra associated with $s_i$ in the generalized framework
of entanglement introduced earlier \cite{Barnum03, Barnum04, Somma}.   
It is also worth mentioning that the same quantity $\eta_{s_i} $, or 
more generally the mean value $\xi_{\cal P}$ in (\ref{xsies}) evaluated for the given partition ${\cal P}$, is related to the quantum Fisher information for the parameter estimation of the low-noise locally depolarizing channels whose actions for quantum states are specified by the partition ${\cal P}$. 
Since the inverse of the quantum Fisher information gives the lower bound of the variance of estimators,
we can provide the operational meaning for $\xi_{\cal P}$ as a measure of precision in the estimation of the strength of the low-noise locally depolarizing channels associated with ${\cal P}$ \cite{BM}. This relation between $\xi_{\cal P}$ and the quantum Fisher information implies that $R_m$ may be interpreted as the quantum Fisher information for an assembly of the low-noise depolarizing channels under the condition that only the number of the local channels is known.

\begin{figure}
\includegraphics[width=2.2in]{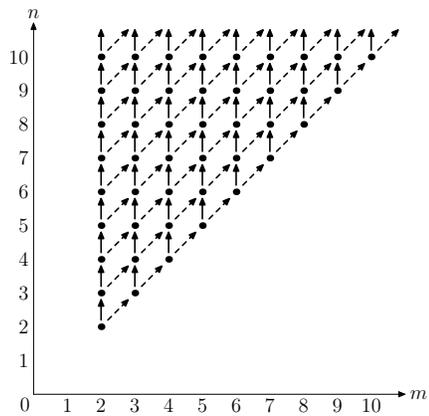}
\caption{The schematic diagram of scalable sequences of $R_m$ for computational complexity. The dot on the lattice point $(m,n)$ represents $R_m$ of $n$-qubit states.  Dotted arrows indicate the polynomial growth required to obtain $R_m$ from the given set of $\eta_{s_i}$, while solid arrows indicate the exponential growth.  The MW measures $R_n$ lie on the lower diagonal edge of the triangle area of the lattice while the measures $R_2$ lie on the vertical left edge.}
\label{CC}
\end{figure}

The computational complexity of $R_m$ may be estimated based on the simple rule that all arithmetic operations (addition, multiplication, division and taking the $k$-th root) are equally counted. 
We then find that, since the number of summations needed for $\rho_{s_i}$ is $2^{n-|s_i|}$, the computational complexity of $R_m$ grows exponentially in general.  Note that this applies to any measures (including the MW measure) which require the partial trace operations.  Further, if $\eta_{s_i}(\psi)$ for all $s_i$ are given, the number of additional steps 
necessary to obtain $R_m(\psi)$ is approximately $m S(n,m)$.
Since $S(n,m)$ grows exponentially for large $n$ with fixed $m$ (see the Appendix), so does  the number of steps for $R_m(\psi)$ from $\eta_{s_i}(\psi)$.  On the other hand, if we are interested in the measures
$R_{n-l}(\psi)$ with fixed $l$ for large $n$ ({\it e.g.}, the MW measure $R_n(\psi)$ arises for $l = 0$), then we see that 
the required number of steps is ${\cal O}(n^{2l+1})$, that is, it grows only polynomially.  These observations show that the computational complexity of $R_m(\psi)$ for $n$-qubit states depends on how to construct the scalable sequences of the measures in question (see Figure \ref{CC}).   The polynomial growth will also arise in general when we restrict ourselves to symmetric states \cite{Stockton}.

The measures $R_m(\psi)$ can readily be extended to those which accommodate mixed states as well \footnote{%
A similar extension of the measure to mixed states is mentioned in \cite{Love} for $m=2$, where 
the convex full extension is carried out at the stage of $\eta_s$ rather than $\xi_{\cal P}$ as we have done here.
However, unlike ours, this does not ensure the equality between $R_m(\rho)=0$ and the corresponding intermediate separability of the state $\rho$ except for $m=2$ which is the case considered in \cite{Love}.
}.
This is done by adopting the standard procedure of considering
the convex hull at the stage of $\xi_{\cal P}$:
\be
\xi_{\cal P}(\rho)=\min_{\{p_\alpha,\psi_\alpha\}}\sum_\alpha p_\alpha\xi_{\cal P}(\psi_\alpha),
\label{genxi}
\ee
where the minimum is chosen from all possible decompositions 
of the density matrix $\rho=\sum_\alpha p_\alpha\ket{\psi_\alpha}\bra{\psi_\alpha}$ into the probability distribution $\{p_\alpha\}$ and the pure states $\ket{\psi_\alpha}$.   From the extended $\xi_{\cal P}(\rho)$ in (\ref{genxi}), we define 
$R_m(\rho)$ as (\ref{rmeasures}).    The resultant measures $R_m(\rho)$ possess the desired property of intermediate 
separability as an extension of (\ref{rmeasurestwo}) to mixed states.  Namely, $R_m(\rho)=0$ iff the state $\rho$ is separable into $m$ subsystems as a mixed state, {\it i.e.}, it admits the form,
\be
\rho =\sum_{\alpha}p_\alpha\bigotimes_{i=1}^m\rho_{s_i}^\alpha,
\label{mixedsep}
\ee
where $\rho_{s_i}^\alpha$ are density matrices in the subsystems $s_i$.   
Note that so defined $R_m(\rho)$ become entanglement measures, since $R_m(\rho)$ are monotone for LOCC and invariant under local unitary operations.

\section{GHZ State vs W state}
We now evaluate the amount of entanglement possessed by the two familiar globally entangled states, 
the GHZ and the W states, using the measures $ R_m(\psi) $ introduced above.  
These are particular states which are invariant under all permutations of constituent subsystems, and this exchange symmetry facilitates our computation considerably.  
To proceed, 
we first note that for those symmetric states the quantity $\eta_{s_i}$ in (\ref{meeta}) depends only on the number of the elements $|s_i|$ of the subset $s_i$, not on the choice of the elements in $s_i$.  
To find the value of the measure $R_m(\psi) $, 
we need to consider all possible partitions ${\cal P}$ with $d({\cal P})=m$ to get the quantity $\xi_{\cal P}(\psi)$
in (\ref{xsies}), but again the exchange symmetry implies that $\xi_{\cal P}(\psi)$ depends only on the way the partition ${\cal P}$ is formed in terms of the set of numbers $|s_i|$ of the elements in the subsets $s_i$ comprising ${\cal P}$. 
To be more explicit, let us choose the numbering of the subsets  $s_i$ in the order $|s_1|\le |s_2| \le \cdots \le |s_m|$ and 
introduce the notation, 
\be
|{\cal P}| :=  \{|s_1|, |s_2|, \ldots, |s_m|\}.
\ee
Note that $|{\cal P}|$ furnishes an ordered partition of the integer $n$ into $d({\cal P}) = m$ nonvanishing integers
by $n = |s_1| + |s_2| + \cdots + |s_m|$.    
For $n$ and $m$ with $2 \le m \le n$, let ${\cal G}(n,m)$ be the set 
of all distinct ordered partitions of the integer $n$ into $m$ nonvanishing integers.   Given some $|{\cal P}| \in {\cal G}(n,m)$,
we denote by $h(|{\cal P}|)$ the total number of partitions ${\cal P}$ sharing the same ordered partition $|{\cal P}|$.
The measure $R_m(\psi)$ in (\ref{rmeasures}) can then be calculated by the product of 
$\xi_{\cal P}(\psi)$
for all different $|{\cal P}|$ in ${\cal G}(n,m)$, {\it i.e.},
\be
 R_m(\psi) = \left(\prod_{|{\cal P}| \in {\cal G}(n,m)}
 \{\xi_{|{\cal P}|} (\psi)\}^{{h}(|{\cal P}| )}\right)^{1/S(n,m)},
\label{rsym}
\ee
where we have written $\xi_{|{\cal P}|}(\psi)$ for $\xi_{\cal P}(\psi)$ to stress that it is dependent only on $|{\cal P}|$.

\begin{figure}
\includegraphics[width=2.55in]{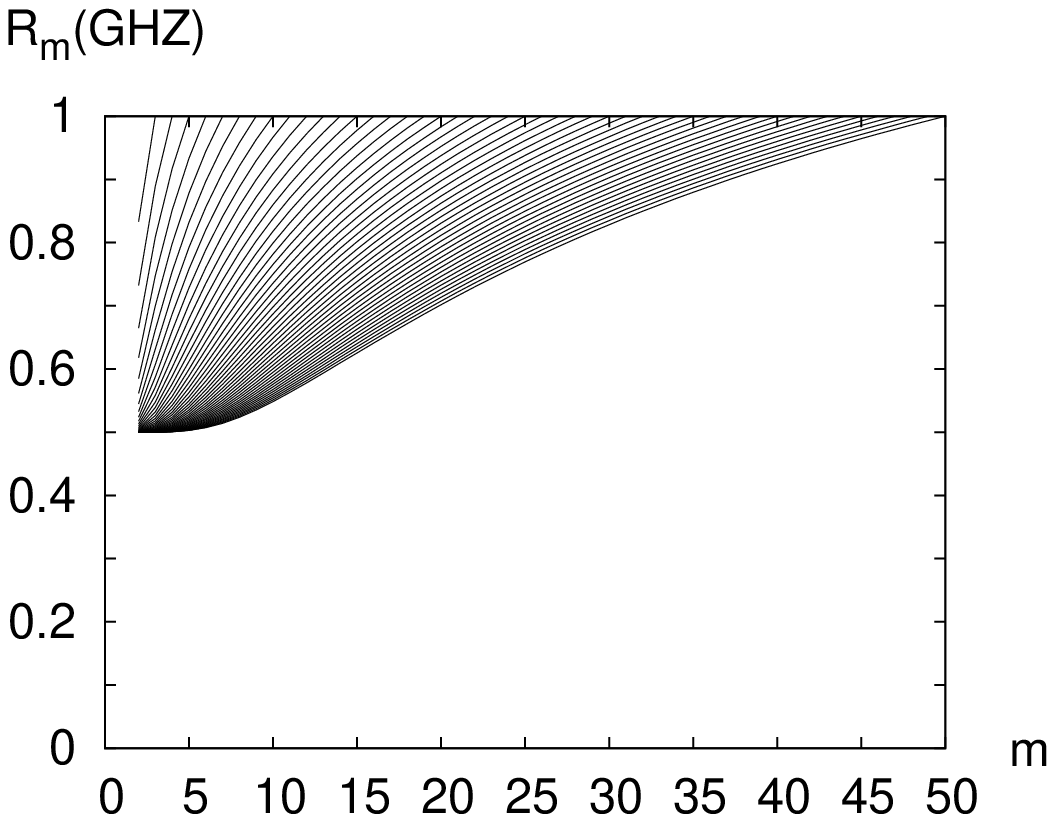}
\includegraphics[width=2.55in]{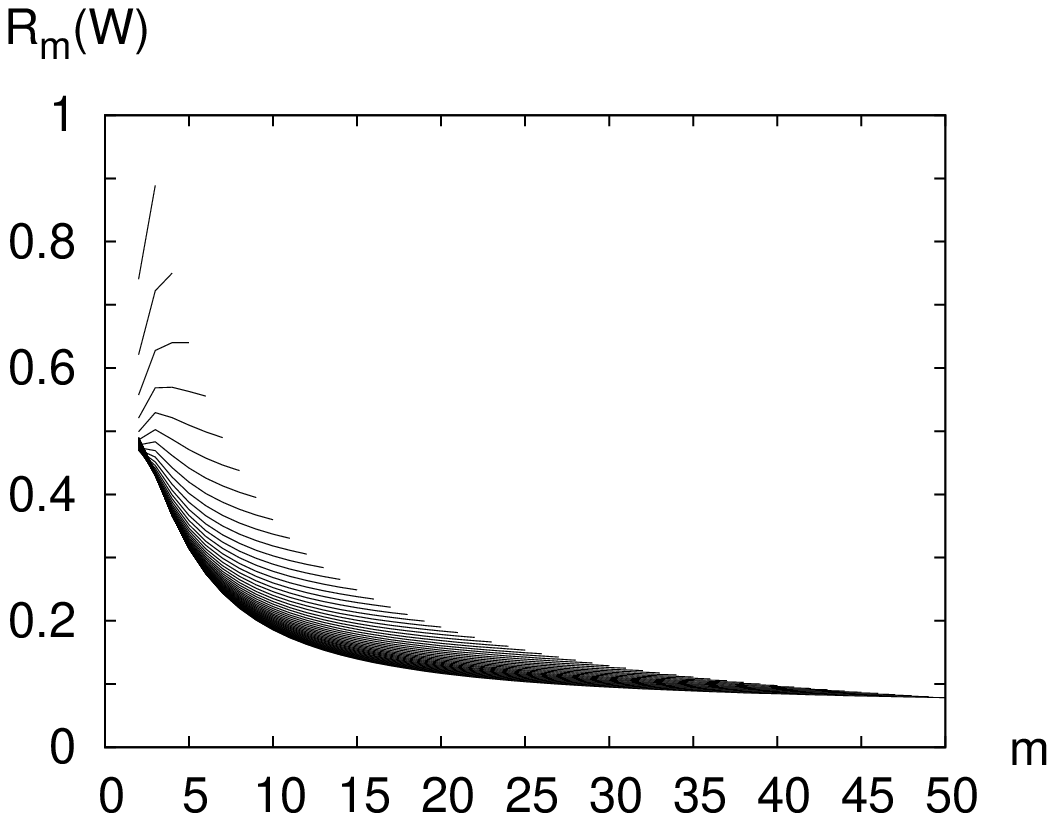}
\caption{The entanglement measures $R_m$ evaluated for the GHZ state $|{\rm GHZ}\rangle$ (above) and the W state $|{\rm W}\rangle$(below) as functions
of $m$ for various $n$ with $3 \le n \le 50$.
Each curve represents $R_m$ for $m$ in the range $2 \le m \le n$ with a fixed value of $n$ which can be read
off from the right end value of $m$ of the curve.}

\label{GHZW}
\end{figure}

Now we consider the $n$-qubit GHZ state (\ref{ghzstate}).
The GHZ state is quite special since it has  ${\rm tr} (\rho_{s_i})^2 = 1/2$ for all subsystems $s_i$, and from this we obtain 
\be
\eta_{s_i}({\rm GHZ}) = \frac{N(|s_i|)}{2}
\label{ghze}
\ee
with $N(|s_i|)$ given in (\ref{meeta}).
To illustrate our procedure for evaluating the measures, we choose, for instance, the case $n=4$, $m=2$ for which  the set ${\cal G}(4,2)$ consists of  the two elements, $|{\cal P}| =  \{1,3\}$ and $\{2,2\}$. 
The numbers of partitions with the same $|{\cal P}|$ are, respectively, 
${h}(\{1,3\})=4!/(1!3!)=4$ and ${h}(\{2,2\})=4!/(2!2!2!)=3$, yielding $S(4,2)=\sum_{|{\cal P}|\in {\cal G}(4,2)}{h}(|{\cal P}|)=7$.  
We then find 
\be
R_2({\rm GHZ})
=\left\{\(\frac{11}{14}\)^4\cdot\(\frac{2}{3}\)^3\right\}^{\frac{1}{7}}
\approx 0.732.
\label{ghzrtwo}
\ee
This procedure can be applied for any $n$ and  $m$, and the results up to $n = 50$ are shown in Figure \ref{GHZW}.

Next, we consider the W state,
\be
|{\rm W}\rangle=\frac{1}{\sqrt{n}}\(\ket{10\cdots 0}+ \ket{01\cdots 0} + \cdots + \ket{00\cdots 1}\),
\ee
which has
\be
\eta_{s_i}({\rm W}) =
N(|s_i|)\frac{2|s_i|(n-|s_i|)}{n^2}.
\label{wet}
\ee
{}For comparison, we again choose the case $n=4$, $m=2$ to find 
\be
R_2({\rm W})
=\left\{\(\frac{33}{56}\)^4\cdot\(\frac{2}{3}\)^3\right\}^{\frac{1}{7}}
\approx 0.621,
\label{wrtwo}
\ee
which is less than the value (\ref{ghzrtwo}) 
of the GHZ state.
As in the GHZ case, the results up to $n = 50$ are shown in Figure \ref{GHZW}.

It is clear from Figure \ref{GHZW} that the GHZ and the W states exhibit rather contrasting behaviors for the entanglement measures $R_m$.   Namely, for the GHZ state, $R_m$ is a monotonically increasing function of $m$ confined within
$1/2 < R_m \le 1$ and approaches the value $R_n = 1$ at the right end $m = n$.  In contrast, for the W state, $R_m$ is basically a decreasing function of $m$ confined in $0 < R_m < 1/2$, except for the small $n < 9$ for which $R_m$ can exceed the value $1/2$.  These can also be seen directly from the formulae (\ref{ghze}) and (\ref{wet}).
In a sense, this result agrees with  our intuitive picture of the GHZ state being more globally entangled than the W state for all $n$, which is also observed by using the entropy of entanglement \cite{Stockton}.
Another point to be noted here is that the clear difference in the values of $R_m$ between the two sets of states suggests that
the GHZ state is more fragile than the W state, because the measures are indirectly related to the fragility of the state which is the source of the operational meaning of preciseness in the estimation discussed in \cite{BM}.  This again is consistent with the conventional view that 
the entanglement of the W state is
more robust than that of the GHZ state \cite{KBI, DVC}.
This propensity of robustness may also be recognized by considering relevant combinations of partitions ${\cal P}$ specific to that purpose as done in \cite{Stockton}.

Note that the lower bound $1/2$ of the measures $R_m$ for the GHZ state indicates that the GHZ state cannot be approximated well by a state which is separable in $m$ subsystems for any number of $m$.  
On the other hand, for the W state we observe that 
the values of $R_m$ with $m$ closer to $n$ approach zero for larger $n$, and in particular, the value $R_n$ ({\it i.e.}, the MW measure) has the vanishing limit,
\be
\lim_{n \to \infty} R_n({\rm W}) = \lim_{n \to \infty}  \frac{n-1}{4n^2}  = 0.
\label{ghzlim}
\ee
This, however, does not mean that the W state $|{\rm W}\rangle$ becomes fully separable in the limit $n \to \infty$.  
One can see this by considering a geometric measure of entanglement $E_G(\psi)$ which is defined by
\be
E_G(\psi):=1-\max_\chi |\langle\chi|\psi\rangle|^2,
\ee
where the maximum is taken over the set of fully separable states $|\chi \rangle$ \cite{Shimony, BL, WG, BDSI}.
Indeed, parameterizing an arbitrary fully separable $n$-qubit state as
\be
|\chi \rangle =\bigotimes_{i=1}^n\left(\cos\theta_i|0\rangle_i+{\rm e}^{{\rm i}\phi_i}\sin\theta_i|1\rangle_i\right),
\ee
and varying the angle parameters in $|\chi \rangle$, one finds that the value of $E_G({\rm W})$ 
is obtained when $\sin^2\theta_i = 1/n$ for all $i$ and $\phi_i=\phi_j$ for all $i, j$.  Hence, in the large $n$ limit we find \cite{BDSI}
\be
\lim_{n\rightarrow\infty}E_G({\rm W})=1-\lim_{n\rightarrow\infty}\left(1-\frac{1}{n}\right)^{n-1}=1-\frac{1}{{\rm e}},
\label{fidW}
\ee
which shows an intriguing fact that despite the vanishing limit of the MW measure $R_n({\rm W})$,
the W state does not approach a definite fully separable state in the limit $n \to \infty$.    This indicates that the connection (\ref{rmeasures}) between the vanishing measure and the separability, which is perfectly valid for finite $n$, does not hold for $n \to \infty$.

\section{Modified W states}
We may further examine the property of our measures by considering a  set of states which are totally symmetric with more than one $\ket{1}$ states in the constituent subsystems.   To be explicit, we introduce the 
\lq ${\rm W}_k$ states\rq\ for 
$1 \le k \le \lfloor n/2 \rfloor$
by 
\be
|{{\rm W}_k}\rangle:= {{n}\choose{k}}^{-\frac{1}{2}}
\left(
\ket{\underbrace{11\cdots1}_{k}\underbrace{00\cdots0}_{n-k}}+\hbox{perm.} \right) ,
\label{wkstate}
\ee
where \lq perm.\rq\ means that all possible distinct terms possessing $k$ \lq1\rq s and $(n-k)$ \lq0\rq s
obtained by permutations of the first term are included.
The ${\rm W}_k$ states, which are known as Dicke states, reduce to the standard W state $|{\rm W}_1\rangle = |{\rm W}\rangle$ for $k=1$, while for $k > 1$ they become slightly more involved but are still manageable thanks to the symmetry.   

\begin{figure}
\includegraphics[width=2.55in]{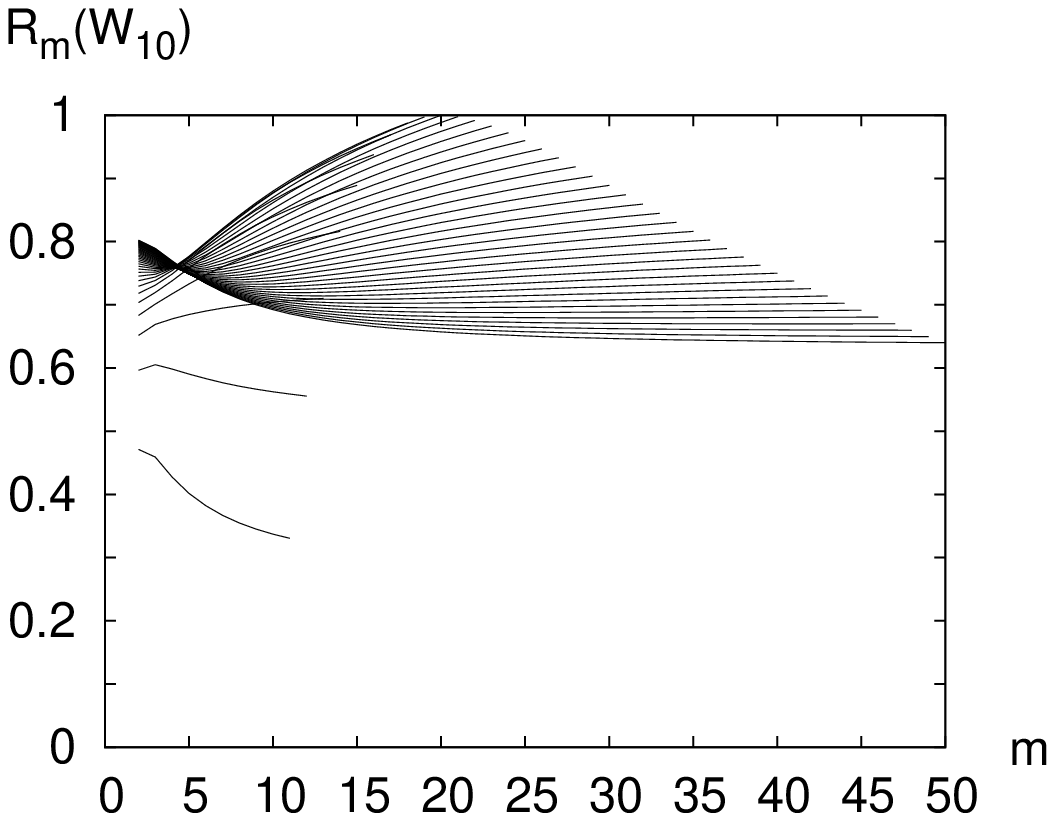}
\includegraphics[width=2.55in]{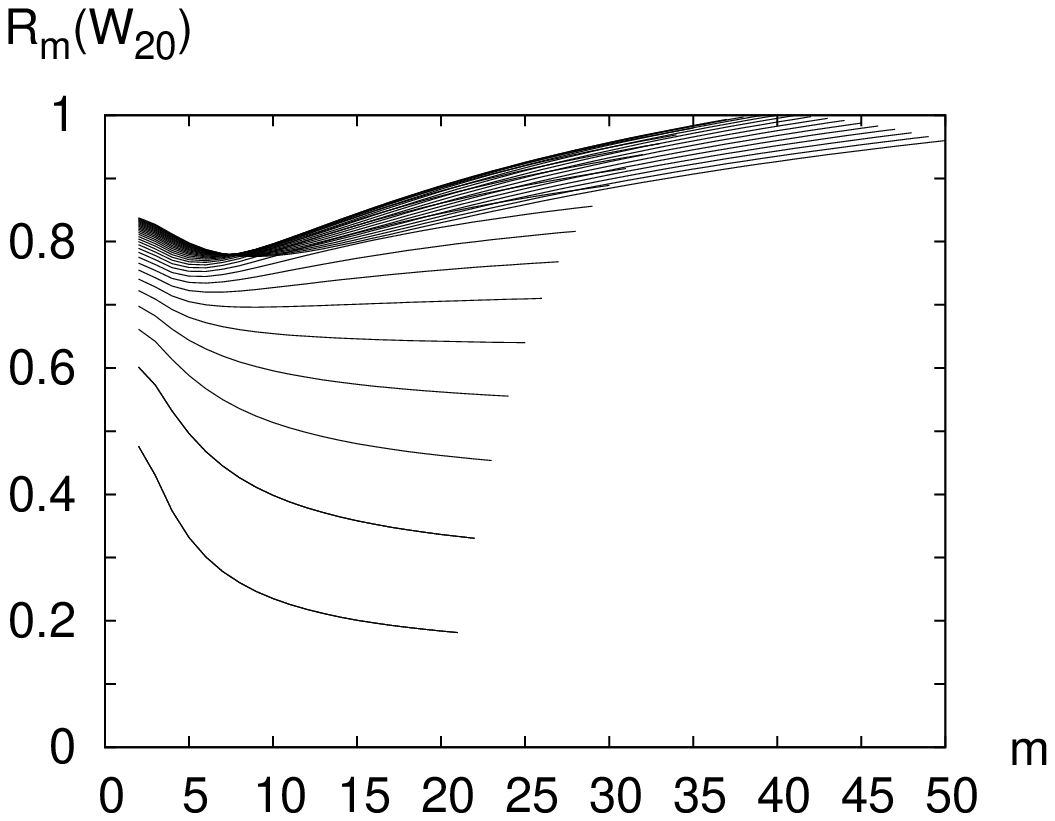}
\caption{The entanglement measures $R_m$ evaluated for the ${\rm W}_{10}$ state (above) and the ${\rm W}_{20}$ state (below) as functions
of $m$ for various $n$ with $k+1 \le n \le 50$.
Each curve represents $R_m$ for $m$ in the range $2 \le m \le n$ with a fixed value of $n$ which can be read
off from the right end value of $m$ of the curve.
}
\label{Wkone}
\end{figure}

To evaluate the measures, we first 
implement an appropriate unitary transformations to $|{\rm W}_k\rangle$ so that, for a given subsystem $s_i$, the state in $s_i$ is represented by the left $|s_i|$ qubits in the $n$-qubit state $\ket{*} = \ket{*}_{s_i} \ket{*}_{\bar s_i}$.    
To proceed, it is also convenient to specify each of the terms in $|{\rm W}_k\rangle$ by the number of \lq1\rq s, which is $k$ for $|{\rm W}_k\rangle$, and an integer $\sigma$ for $1 \le \sigma \le {{n}\choose{k}}$ labeling the distinct terms appearing in the permutations.  Clearly, the same notation can be employed for both of the subsystems $s_i$ and $\bar s_i$ as well, and we may write
an arbitrary term in (\ref{wkstate}) as a product of states in the two subsystems as
$\ket{k, \sigma} = \ket{r, \tau}_{s_i} \ket{k-r, \tau'}_{\bar s_i}$, 
where $\ket{r, \tau}_{s_i}$ is a state of the subsystem $s_i$ with $r$ \lq1\rq s and the label $\tau$ runs for $1 \le \tau \le {{\vert s_i\vert}\choose{r}}$, and similarly $\ket{k-r, \tau'}_{\bar s_i}$ is a state of the subsystem $\bar s_i$ with the label $\tau'$ running over $1 \le \tau' \le {{n - \vert s_i\vert}\choose{k-r}}$.
This allows us to rewrite the ${\rm W}_k$ state (\ref{wkstate}) in the form,
\be
|{\rm W}_k\rangle = {{n}\choose{k}}^{-\frac{1}{2}}
\sum_r
\sum_{\tau,\tau'} \ket{r, \tau}_{s_i} \ket{k-r, \tau'}_{\bar s_i}
\ee
from which the reduced density matrix is found as
\be
\rho_{s_i} = {{n}\choose{k}}^{-1}
\sum_r
{{n - \vert s_i\vert}\choose{k-r}}
\sum_{\tau,\tau'} \ket{r, \tau}_{s_i} {}_{s_i}\!\bra{r, \tau'},
\ee
where the summation of $r$ is for $\max(|s_i|-(n-k),0) \le r \le  \min(|s_i|,k)$.
It is now straightforward to evaluate $\eta_{s_i}$ to find
\be
\eta_{s_i}({\rm W}_k ) = N
\left\{1 - {{n}\choose{k}}^{-2}
\sum_r
{{\vert s_i\vert}\choose{r}}^2{{n - \vert s_i\vert}\choose{k-r}}^2\right\}\!\! .\,\, \,\, 
\label{wketa}
\ee

\begin{figure}
\includegraphics[width=2.55in]{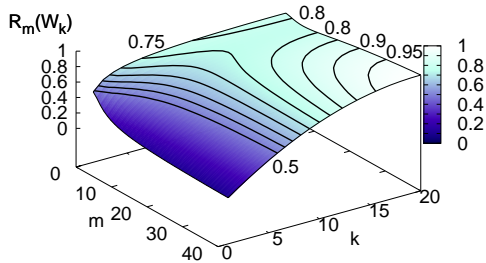}
\caption{(Color online) The entanglement measures $R_m$  evaluated for the ${\rm W}_k$ states as functions
of $m$ and $k$ under the fixed value of $n = 40$.}
\label{Wktwo}
\end{figure}

Based on the result (\ref{wketa}), one can obtain the values of the measures $R_m$ for the ${\rm W}_k$ states, and the outcomes are shown 
in Figure \ref{Wkone} for the two cases $k ={10}$ and $k = {20}$.   
It is seen that both of the ${\rm W}_{10}$ and ${\rm W}_{20}$ states exhibit distinctive behaviors which are also different from those of the GHZ states and the W states discussed before, and in particular we notice that the measure $R_m$ achieves the upper limit $R_m = 1$ at $n=m=2k$.  
This can also be confirmed from (\ref{wketa}) since 
for symmetric states, $n=m$ implies $R_m(\psi)= \eta_{s_i}(\psi)$ with $\vert s_i\vert=1$, which is 1 for $n=2k$.
We therefore see that the $n$-qubit ${\rm W}_{n/2}$ state furnishes the maximally entangled state for the MW measure.
This may be understood from the large symmetry possessed by the ${\rm W}_{n/2}$ state which has an equal number of $\ket{0}$ and $\ket{1}$ states in the constituent subsystems.

It is also interesting to look at the behaviors of the $R_m$ measures for the ${\rm W}_k$ states as functions of $k$ with some fixed $n$.   This can be done in Figure \ref{Wktwo}, where we plot the values of $R_m$ for ${\rm W}_k$ in the range 
$2\le m \le 40$ and $1 \le k \le 20$ for $n = 40$.   We
observe there that $R_m$ are monotonically increasing functions of $k$ for all $m$, indicating that the ${\rm W}_k$ states are \lq more entangled\rq\ for larger $k \le n/2$ under all measures $R_m$.    
Moreover, we see that the change in the values of the measure is in general more prominent for $R_m$ with higher $m$, which suggests that  variation of $k$ alters the ${\rm W}_k$ states in their entanglement property of higher separability.

\section{Conclusions and Discussions}

In this paper we have introduced a family of entanglement measures $R_m$, $m = 2, \ldots, n$, for $n$-qubit states in which both the MW measure \cite{MW} and the measure proposed by Love {\it et al.}~\cite{Love} arise as the two extreme cases $m = n$ and $m = 2$.  Our measures $R_m$ are scalable and can be used as indicators for separability of the $n$-qubit states into $m$ subsystems. 
We have seen by comparing the behaviors of the GHZ state and the W state that the measures $R_m$ are also useful to characterize the entangled nature even for globally entangled states.   The ${\rm W}_k$ states, which give the maximally entangled state for the MW measure at $k = n/2$,  furnish another set of globally entangled states which behave distinctively under  $R_m$ depending on $k$.

The outcomes of the analysis on the GHZ state and the W state indicate that, in terms of our measures $R_m$, the GHZ state is more entangled than the W state.  This is to be contrasted to the observation in \cite{WG, BDSI} that, in terms of the geometric measures, the converse holds.  
This may be derived from the difference in the basic ingredient of the measures, that is, for bipartite systems our measures $R_m$ are a generalization of the linear entropy, whereas the geometric measures are related to Chebyshev entropy \cite{BZ}.    Apart from the origin of the difference, this poses 
the question on the relation between the two families of entanglement measures, 
and calls for clarification of the physical and operational meanings of these measures other than the aforementioned preciseness of estimation valid for the constituent element $\xi_{\cal P}$.  
Along with the formal classification of our measures with respect to the generalized framework of measures \cite{Barnum03, Barnum04, Somma}, 
these issues should be investigated further to gain a fuller picture of multipartite quantum entanglement in general.

\begin{acknowledgments}
This work has been supported in part by
the Grant-in-Aid for Scientific Research (C), No.~20540391-H20, and
Global COE Program \lq\lq the Physical Sciences Frontier\rq\rq, MEXT, Japan.
\end{acknowledgments}

 \renewcommand{\theequation}{A.\arabic{equation}}
 \setcounter{equation}{0}  
 \section*{APPENDIX}  

This appendix provides an elementary account of 
the asymptotic behavior of the Stirling number in the second kind $S(n,m)$ for large $n$, which is needed to 
estimate the computational complexity of $R_m(\psi)$ in the text.   Recall first that $S(n,m)$ fulfills the recurrence relation,
\be
S(n,m)-S(n-1,m-1)=mS(n-1,m)
\label{rec}
\ee
with the initial conditions $S(n,n)=S(n,1)=1$.   From this we immediately obtain
\be
S(n,m)>mS(n-1,m)>\cdots>m^{n-m},
\ee
which shows that, for fixed $m$, the number of steps to compute $R_m(\psi)$ grows exponentially as ${\cal O}(m^n)$. 

Next, we consider the case where $n$ increases with fixed $l$ as discussed in Sec.~II.  For example, if $l=2$, 
we sum up the recurrence relations (\ref{rec}) with $n = k$, $m = k-2$ from $k=4$ to $k=n$ and use 
$S(n,n-1)={n\choose2}$ to obtain
 \be
\!\!\! \!\!\! S(n,n-2)=\frac{1}{4!}n(n-1)(n-2)(3n-5)={\cal O}(n^4).
 \ee
More generally, if we assume $S(n,n-(l-1))={\cal O}(n^{2(l-1)})$, then by the same iterative procedure for (\ref{rec}), we obtain $S(n,n-l)={\cal O}(n^{2l})$.   By induction, this gives the polynomial growth of $S(n,n-l)$ for large $n$.

With the help of Stirling\rq s approximation, a more explicit asymptotic formulas of $S(n,m)$ can be found.   Namely, for $n\gg m$, $S(n,m)$ is written as \cite{BW74}
\begin{equation}
S(n,m)\approx \frac{1}{\sqrt{2\pi}}\frac{m^n}{m!},
\end{equation}
which shows the exponential growth of $S(n,m)$ for large $n$
and is consistent with our result $S(n,m)=\mathcal{O}(m^n)$.
In \cite{Hsu}, we also find
\begin{equation}
S(n,m)\approx \frac{\(\frac{1}{2}m^2\)^{n-m}}{(n-m)!} 
\end{equation}
for large $n$ and small $n-m$, which confirms our result $S(n,n-l)={\cal O}(n^{2l})$.


\end{document}